\documentclass[conference]{IEEEtran}
\usepackage{psfig,amssymb,amsmath,amsfonts,amsthm}

\newtheorem{theorem}{\textbf{Theorem}}
\newtheorem{lemma}{\textbf{lemma}}

\title{Collaborative Downloading in VANET using Network Coding}

\begin{document}
\author{Mohammad~H.~Firooz,~\emph{Student~Member,~IEEE,}
        Sumit~Roy,~\emph{Fellow,~IEEE,}\\
        Electrical Engineering Department\\
        University of Washington\\
        Seattle, WA 98105.\\
        \{firooz,sroy\}@u.washington.edu
       }
\maketitle
\begin{abstract}
Data downloading on the fly is the base of commercial data services in vehicular
networks, such as office-on-wheels and entertainment-on-wheels. Due to the sparse spacial distribution
of roadside Base Stations (BS) along the road, downloading through Roadside-to-Vehicle (R2V) connections is intermittent. When
multiple vehicles with geographical proximity have common interest in certain objects to download, they can collaborate to reduce significantly their overall download time. In this paper, we investigate application
of Network Coding (NC) in collaborative downloading (CD). We focus on the R2V part of CD, and analytically derive probability distribution and expected value of amount of time needed to deliver all information to the vehicles with and without NC. Our results show that using NC slightly improves the downloading time in addition to removing any need for having any sort of uplink communications from vehicles to the infrastructure.
\end{abstract}
\section{Introduction}
Recent development and standardization
in vehicular ad hoc networks (VANETs) \cite{hartenstein2009vanet} have motivated
increasing interest in data services for in-vehicle
consumption, such as 'commerce- and entertainment-on-the-wheel' \cite{IEEEWave}.
Hence, in near future, the number of vehicles equipped with wireless communication devices is poised to increase dramatically, i.e., we are moving closer to an intelligent transportation system.
Such system would provide a wide variety of applications: local
information pushed to vehicles (e.g., traffic notification, map updates, location based
advertisements); or specific data
pulled from Internet servers (e.g., neighborhood parking, reviews
of local restaurants, and video clips of local attractions) \cite{zhu2003mac}.

In current intelligent transportation system, services are provided to vehicles using existing wide-area cellular infrastructure (3G/4G) and/or
the roadside infrastructure based on
Dedicated Short Range Communications (DSRC) links\footnote{DSRC in North America is based on 802.11p wireless link access in 5.9 GHz band, subsequently folded into the IEEE 1609 Wireless Access for Vehicular Environments (WAVE) standards.} proposed for VANET networks \cite{IEEEWave}. However, both of these approaches have their own challenges:
the modest data rate of present 3G links and its high cost of
data downloading on one hand, and the intermittent hotspot
type roadside coverage envisaged with DSRC on the other
\cite{jiang2008ieee}.

The above challenges lead to the following simple premise for content dissemination, called collaborative downloading:
if the content is available at a subset of vehicles
that is also desired by
(many) others in the network, peer-to-peer content distribution
using vehicle-to-vehicle (V2V) ad hoc communications, is
time and cost efficient.

Collaborative downloading is a data dissemination protocol, distributing information among all nodes inside the network \cite{korkmaz2004urban},
\cite{xu2004opportunistic}, and it has attracted a lot of attention in VANET community in the past few years \cite{ghandeharizade2004pavan}\cite{Fei}. In general,
data dissemination in VANET networks consists of two phases \cite{jerbi2007experimental}\cite{jerbi2008characterizing}:
\begin{enumerate}
  \item Roadside-to-Vehicle (R2V) phase: in this phase vehicles are communicating with a base station (located in specific location on the road) to receive data.
  \item Vehicle-to-Vehicle (V2V) phase: when the vehicles are out of the coverage of the BS, they try to exchange information between each other. If this phase is completed all the nodes have the same data.
\end{enumerate}

\begin{figure}
\centering
\begin{tabular}{c}
\psfig{figure=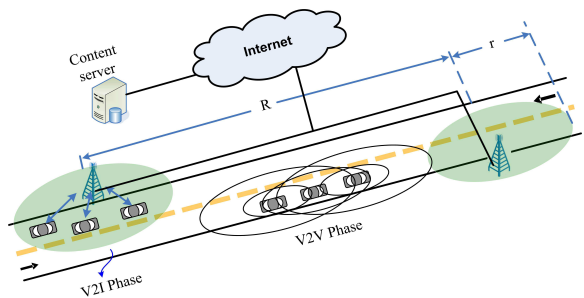,width=3.0in}
\end{tabular}
\caption{System diagram of collaborative downloading among vehicles on the road.}
\label{fig:system}
\end{figure}

To better understand the concept of collaborative downloading, consider the scenario in Figure \ref{fig:system} in which a group of vehicles equipped with a DSRC radio are connected to the Internet via roadside DSRC base stations. Assume that these vehicles have common interest in a large file on the Internet. The intermittency arises from the sparsity of the roadside DSRC base-stations and the hotspot nature of coverage. As a result, data is downloaded during the (short) intervals of radio connectivity from the infrastructure to the vehicle.  Thus each vehicle in a platoon only obtains a fraction of the overall file in each contact duration.
In principle, each vehicle could download the whole file through multiple contacts, requiring significant latency due
the sparsity of the contacts. A more effective alternative is for the vehicles to form a coalition for sharing their respective pieces after each round of contacts (collaborative downloading) when they are out of coverage (V2V phase).

Our focus in this paper is on R2V phase. We assume a perfect V2V transmission, i.e., data dissemination is complete after V2V phase. In other words, we assume that the vehicles have the same information after V2V phase is over.

In the state-of-the-art algorithm for collaborative downloading, nodes need to communicate with BS to let it know which packets they are interested in. That means, a partial time of vehicle to BS connection (in R2V phase) has to be assigned to signalling from the vehicles to the BS\footnote{RVC links are half-duplex, i.e., the base station can not receive and transmit simultaneously.}. Moreover, this scheme requires a lot of synchronization and handshaking between the BS and the receiving BS. In this paper, we propose to use network coding in downloading data from the infrastructure. Our algorithm omits any kind of signaling from vehicles to the BS. Moreover, we analytically show that expected time needed to disseminate data from infrastructure to all vehicles is slightly less by using NC. In other words, NC slightly decreases the downloading delay in addition to removing any need for uplink communication.

The rest of the paper is organized as follows: In section \ref{S:review}, we briefly review application of NC in wireless networks. Section \ref{S:I2V_mode} present our system model for collaborative downloading. We derive dissemination delay when NC is not exploited in Section \ref{S:DissNoNC}. Average amount of time needed to deliver information to all nodes in the network with use of NC is derived in Section \ref{S:DissNC}. Our evaluation result is given in Section \ref{S:Eval}. The paper concludes with reflections on future works in Section \ref{S:conclude}. The proofs of theorems is given in the Appendix.

\emph{Notations}: We use bold capitals (e.g. $\mathbf{A}$) to represent matrices and bold lowercase symbols (e.g. $\mathbf{m}$) for vectors. The $i$-th entry of a vector $\mathbf{m}$ is denoted by $m_i$ and Superscript $^T$ denotes matrix transpose. We use calligraphic capitalized symbols (e.g. $\mathcal{T}$) for random variables.

\section{Data Dissemination in Wireless Networks Using Network Coding: A Review}\label{S:review}
Network coding has received considerable attention in recent
years for its potential for achieving the theoretical upper
bound (max-flow, min-cut) of network resource utilization via
the introduction of coding concepts at the network layer \cite{koetter2003aan,li2003lnc}. It has
been shown that with simple random linear coding in place of
the usual forwarding, system throughput can be increased in
several canonical network topologies \cite{ho2003rnc,ho2006rln}.
Although network coding was developed in the context of efficient multicasting \cite{ahlswede2000network}, it has since been adapted to an increasing set of new applications \cite{hamedNC,ho2005nmm,gkantsidis2006cooperative}. Recently, random linear network coding within a gossip-based dissemination protocol was proposed in \cite{deb2006algebraic} for wired networks. They show that in a fully-connected network, information dissemination schemes based on network coding can provide substantial benefits. \cite{mosk2006information} extends the result in \cite{deb2006algebraic} by deriving (loose) upper bounds, for a wired network with arbitrary topology.

Increased throughput resulting from use of network coding is not limited to wired networks. In \cite{katti2008xors}, authors apply network coding to unicast flow in wireless networks. Their results show that significant gain (in terms of throughput) can be achieved even in the case of unicast, using simple XOR combining. Sagduyu et al. investigate the interaction between MAC and network coding, devising suitable conflict-free transmission schedules in wireless multi-hop network \cite{evren2007joint}. In \cite{yomo2007opportunistic}, Yomo and Popovski propose distributed and opportunistic scheduling rules for combining packets in the presence of time-varying fading channel.

Recently network coding has been applied to the data dissemination problem in wireless networks. In \cite{fragouli2008efficient}, authors use gossip based algorithm (called rumor-spreading) to diffuse information in an ad-hoc network. Their work is continued by \cite{asterjadhi2010toward} who study the performance degradation due to actual MAC schemes.

\section{Roadside-to-Vehicle Phase:System Model}\label{S:I2V_mode}

Consider a platoon of $N$ vehicles interested in a certain file comprising of $M$ packets $x_1, x_2,\ldots x_M$. Suppose the network infrastructure (i.e., all the roadside base stations) possesses that common file. The goal is to distribute this file to all vehicles in the network. This is achieved by cycling through a succession of R2V+V2V phases. During each R2V phase, we assume that any of the $N$ vehicles downloads a constant number of $m\ll M$ packets, i.e., the duration and rate of download per contact is constant and independent of the vehicle. For security assurance, we assume that communications between the BS and each vehicle is encrypted. In other words, the BS communicates with only one vehicle at a time and the other nodes can not see the communication link between them. That means, in collaborative downloading sharing is limited to V2V phase and we do not use the advantage of broadcasting in R2V phase for the sake of security.

Obviously, the way the $m$ packets are chosen affects the system performance. We consider the two following ways of selecting $m$ packets in each round:
\begin{itemize}
  \item \emph{Feedback-based scheme}: The $m$ packets are chosen randomly by the serving BS among the \emph{currently unreceived packets} of the vehicle. It is assumed that in each round, nodes can individually signal to the server the specific indices of packets yet to be received.
  \item \emph{Network coding aided scheme}: In this scheme BS uses Random Linear Network Coding (RLNC). In each transmission, it sends a linear combination of the $M$ packets, where the combining coefficients are uniformly chosen over the finite field $\mathbb{F}_{2^q}$. In this scenario no feedback is needed from vehicles to the BS.
\end{itemize}

%SR The algorithm for RNC is not clear. Are the RNC coeff. a) indep. chosen between vehicles in each round and b) indep. chosen between
%SR rounds? Also mention that there is no feedback needed from vehicle to BS (i.e. BS chooses the RNC coeff. randomly without any
%SR inputs from the vehicles (unlike the other scheme). Basically, each vehicle builds up it's A matrix till it hits full rank and
%SR the algorithm ends.

%mhf As it is mentioned, in each round, coefficients are chosen uniformly random in \mathbb{F}_{2^q} no extra condition
%is needed.

When the last vehicle leaves the BS coverage area, all nodes are in V2V phase. In this phase, every node tries to share its information with other nodes inside the network. After V2V phase is complete, all vehicles have the same information. When the vehicles enter the range of the next BS, they try to obtain the remaining missing packets. This continues until every node has the full message of $M$ packets. We call each R2V and its following V2V phase a \emph{round}. The number of rounds required to send the $M$ packets from the infrastructure to the vehicles is the matter of interest. In this paper, we aim to derive probability distribution and expected value of the number of rounds needed to disseminate information to all $N$ vehicles in the network, using either the feedback-based or NC-aided scheme.

For the sake of simplicity, we only consider the case where there are two vehicles in the network, i.e., $N=2$. Generalizing our result to arbitrary $N$ is considered as one of our future works. For each scheme, we first analyze a simple case of $m=1$, transmitting only one new packet in each R2V phase of a vehicle. Then, we generalize our result to arbitrary $m$.
\begin{figure}
\psfig{figure=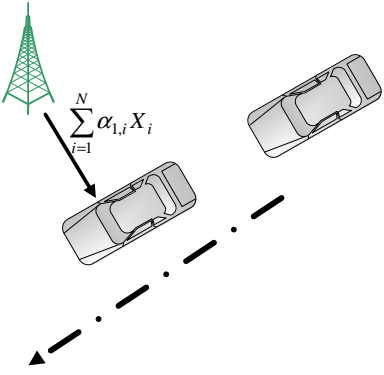,width=2.0in}
\caption{In each transmission, the BS sends a linear combination of the $M$ packets to a vehicle and this transmission can not be seen by any other vehicles.}
\label{F:I2V_simple}
\end{figure}
\section{Dissemination Without Network Coding}\label{S:DissNoNC}
Suppose there are $N=2$ vehicles in the network. As mentioned before, in each R2V phase, a vehicle downloads $m$ packets from the BS. In the following V2V phase, vehicles exchange their uncommon packets. In the $i$-th round, let $\mathcal{X}_i$ denote the number of \emph{common} packets among the $m$ downloaded packets by each node. Clearly, $\mathcal{X}_i$ is a random variable in the set $\{0,1,\ldots,m\}$. Further, let define $\mathcal{S}_i$ as number of packets each node has at the end of $i$-th round. Clearly, at the end of each round (R2V+V2V) $2m-\mathcal{X}_i$ new packets are added to each node. Thus, It is easy to see the following recursive equation for $\mathcal{S}_i$:

\begin{equation}\label{E:recursiveS_i}
\mathcal{S}_i = \mathcal{S}_{i-1}+2m-\mathcal{X}_i.
\end{equation}

The dissemination algorithm stops when both cars have all the $M$ packets, i.e., the whole information. Let $\mathcal{T}$ denote stopping time defined as follows:
\begin{equation}
\mathcal{T}=\min_t\{\mathcal{S}_t = M, t\ge 0\}.
\end{equation}

In this section we aim to calculate probability distribution and expected value of the stopping time for the feedback-based dissemination algorithm.

\subsection{$m=1$, $N=2$}

\begin{lemma}\label{Tm:P(X_i)}
The number of common packets at the i-th round given $\mathcal{S}_{i-1}=s$ has a probability distribution given by
\begin{eqnarray}
P(\mathcal{X}_i=x|\mathcal{S}_{i-1} = s_i)  &=& {{1\choose x}{M-s_i-1\choose 1-x}\over {M-s_i \choose 1}} \\ \nonumber
                        &=& {{M-s_i-1\choose 1-x}\over {M-s_i}} \,\,\,x=0,1.
\end{eqnarray}
\end{lemma}
\emph{Proof:} see the appendix
\begin{flushright}
$\blacktriangle$
\end{flushright}

As mentioned before, a vehicle, in each R2V phase, individually signal to the BS the specific indices of packet yet to be received by it. Thus, in each round number of packets each node has, would be increased by at least one when $m=1$. Hence number of
rounds needed to deliver all the $M$ packets is at most $M$, i.e., $\mathcal{T}$ is bounded from above by $M$. On the other hand, at
each round at most two new packets can be delivered to the vehicles. Therefore, stopping time is bounded from below by $M/2$.
By the same argument, it is easy to see that for each $0\le t\le \mathcal{T}$, $\mathcal{S}_t$ is bounded as $t\le \mathcal{S}_t\le 2t$.

%SR Why is there an upper bound on \mathcal{T}? If both cars keep getting the same packet in successive rounds, then \mathcal{S}_i stays the same; there
%SR is no upper bound on \mathcal{T}. Min. value of \mathcal{T} = M/2 is OK.

%mhf reply: I explained it.

By definition of $\mathcal{T}$, we have $\mathcal{S}_\mathcal{T}=N$. We aim to calculate probability distribution of stopping time, i.e., $P(\mathcal{T}=t)$.
If the stopping time is $\mathcal{T}$, then at $\mathcal{T}-1$, one and only one of the followings is correct:
\begin{itemize}
  \item $\mathcal{S}_{\mathcal{T}-1}=M-1$
  \item $\mathcal{S}_{\mathcal{T}-1}=M-2$
\end{itemize}

By conditioning on the above events and by using the law of total probability, $P(\mathcal{T}=t)$ can be calculated as follows:

\begin{eqnarray}\label{E:P(T=t)}
P(\mathcal{T}=t) &=& P(\mathcal{T} = t|\mathcal{S}_{t-1} = M-1)P(\mathcal{S}_{t-1} = M-1) \\ \nonumber
&&+P(\mathcal{T} = t|\mathcal{S}_{t-1} = M-2)P(\mathcal{S}_{t-1} = M-2) \\ \nonumber
          &=& P(\mathcal{S}_t = M|\mathcal{S}_{t-1} = M-1)P(\mathcal{S}_{t-1} = M-1) \\ \nonumber
          &&+ P(\mathcal{S}_t = M|\mathcal{S}_{t-1} = M-2)P(\mathcal{S}_{t-1} = M-2).
\end{eqnarray}

Given $\mathcal{S}_{t-1}=M-1$, probability of having $\mathcal{S}_t=M$ is the same as probability of $\mathcal{X}_t=1$, which has been given in Eq. \eqref{E:recursiveS_i}. Thus, by using Lemma \ref{Tm:P(X_i)}, we have:

\begin{eqnarray*}
P(\mathcal{S}_t = M|\mathcal{S}_{t-1} = M-1)  &=&   P(\mathcal{X}_t=1|\mathcal{S}_{t-1} = M-1)  \\ \nonumber
&=& 1,
\end{eqnarray*}
 because if $\mathcal{S}_{t-1} = M-1$, it means that the vehicles are one packet short. In the next round, the BS will send them that packet to complete the downloading process.

The above can be summarized to the following lemma:

%SR What does $t \le \mathcal{S}_t\le 2t$ mean? t is a constant, you need to use \mathcal{S}_\mathcal{T} to denote any event based on stopping time \mathcal{T}.

%mhf reply: I talked about $t \le \mathcal{S}_t\le 2t$ in previous paragraph. Also \mathcal{S}_t means we are at state (or cycle)
%t and we know \mathcal{S}_\mathcal{T}=M (by definition) which means at the last cycle (when the algorithm stops) we have M packets at both cars

\begin{lemma}
Let $\mathcal{S}_t$ be the number of packets each node has at the end of t-th round. Then
\begin{eqnarray}
P(\mathcal{S}_t = s+1|\mathcal{S}_{t-1} = s) &=&  P(\mathcal{X}_t=1|\mathcal{S}_{t-1} = s)  \\ \nonumber
                         &=&  {1\over {M-s}},
\end{eqnarray}

and

\begin{eqnarray}
P(\mathcal{S}_t = s+2|\mathcal{S}_{t-1} = s) &=&  P(\mathcal{X}_t=0|\mathcal{S}_{t-1} = s)  \\ \nonumber
                         &=&  {{M-s-1}\over{M-s}},
\end{eqnarray}
where (as mentioned before)
\begin{equation}
t \le \mathcal{S}_t\le 2t.  \nonumber
\end{equation}
\end{lemma}
\begin{flushright}
$\blacktriangle$
\end{flushright}

To calculate $P(\mathcal{T}=t)$ in Eq. \eqref{E:P(T=t)}, we need to calculate $P(\mathcal{S}_t = s)$. It can be calculated by conditioning
on the two following collectively exclusive events:
%SR Don't understand this, \mathcal{T} is a stopping time, so $\mathcal{T}=t$ is equivalent to $\mathcal{S}_t = M$. $\mathcal{S}_t= s $ is unrelated to stopping time.

%mhf reply: no, that's not true for example if \mathcal{S}_t0 = M-1. then you know for sure that \mathcal{T}=t0+1. In another example if
%\mathcal{S}_t1=M-3 you know that \mathcal{T} is either t1+2 or t1+3.

\begin{itemize}
  \item $\mathcal{S}_{t-1} = s-1$
  \item $\mathcal{S}_{t-2} = s-2$
\end{itemize}

Therefor, we have
\begin{eqnarray}
P(\mathcal{S}_t = s)\!\! &=& \!\!P(\mathcal{S}_t = s|\mathcal{S}_{t-1} = s-1)P(\mathcal{S}_{t-1}=s-1) \\ \nonumber
&&+P(\mathcal{S}_t = s|\mathcal{S}_{t-1} = s-2)P(\mathcal{S}_{t-1}=s-2) \\ \nonumber
           \!\!&=& \!\!P(\mathcal{X}_t = 1|\mathcal{S}_{t-1} = s-1)P(\mathcal{S}_{t-1}=s-1) \\ \nonumber
           &&+P(\mathcal{X}_t = 0|\mathcal{S}_{t-1} = s-2)P(\mathcal{S}_{t-1}=s-2) \\ \nonumber
           \!\!&=&\!\! {1\over{M-s+1}}P(\mathcal{S}_{t-1}=s-1) \\ \nonumber
           &&+ {{M-s+1}\over {M-s+2}}P(\mathcal{S}_{t-1}=s-2).
\end{eqnarray}

So, $P(\mathcal{S}_t = s)$ can be recursively calculated using above equation and the following initialization:

\begin{eqnarray}
P(\mathcal{S}_1 = 1) &=& P(\mathcal{X}_1 = 1) = {1\over M}, \\ \nonumber
P(\mathcal{S}_1 = 2) &=& P(\mathcal{X}_1 = 0) = {{M-1}\over M}. \\ \nonumber
\end{eqnarray}

We summarize the above findings to the following theorem:

\begin{theorem}
Let $\mathcal{T}$ be the stopping time. Then, it has a probability distribution as follows:
\begin{equation}
P(\mathcal{T}=t) = P(\mathcal{S}_{t-1} = M-1) + {1\over 2}P(\mathcal{S}_{t-1} = M-2),
\end{equation}
where ${M\over 2}\le t \le M$. $P(\mathcal{S}_t=s)$ can be calculated using the following recursive formulation:
\begin{eqnarray}
P(\mathcal{S}_t = s) &=& {1\over{M-s+1}}P(\mathcal{S}_{t-1}=s-1) \\ \nonumber
&&+ {{M-s+1}\over {M-s+2}}P(\mathcal{S}_{t-1}=s-2),
\end{eqnarray}
which has the following initialization expressions:
\begin{eqnarray}
P(\mathcal{S}_1 = 1) &=& {1\over M}, \\ \nonumber
P(\mathcal{S}_1 = 2) &=& {{M-1}\over M}. \\ \nonumber
\end{eqnarray}
\end{theorem}
\begin{flushright}
$\blacktriangle$
\end{flushright}

\subsection{$N=2$ and arbitrary $m$}

The result in previous section can be readily generalized to the following. Its proof is omitted here for the sake of space limitation.
\begin{theorem}
Let BS have $M$ packets to distribute among two vehicles. In each R2V phase it transmits $m$ packets to each car separately. Let $\mathcal{T}$ denote the stopping time. Then, it has the following probability distribution:
\begin{equation}
P(\mathcal{T}=t) = \sum_{i=1}^{2m} {{m\choose 2m-i}\over {i\choose m}}P(\mathcal{S}_{t-1}=M-i),
\end{equation}
where ${M\over {2m}}\le t\le {M\over m}$. Moreover, $P(\mathcal{S}_t=s)$ can be calculated using the following recursive formula:
\begin{equation}
P(\mathcal{S}_t=s)=\sum_{i=m}^{i=2m}{{{m\choose 2m-i}{M-s+i-m\choose i-m}}\over {M-s+i\choose m}}P(\mathcal{S}_{t-1}=s-i),
\end{equation}
which has initialization expressions as follows:
\begin{equation}
P(\mathcal{S}_1 = i) = {{{m\choose 2m-i}{M-m\choose i-m}}\over {M\choose m}},\,\,\, i=m,\ldots,2m.
\end{equation}
\end{theorem}
\emph{Proof}: see \cite{hamed}
\begin{flushright}
$\blacktriangle$
\end{flushright}

%%%%%%%%%%%%%%%%%%%%%%%%%%%%%%%%%%%%%%%%%%%%%%%%%%%%%%%%%%%%%%%%%%%%%%%%%%%%%%%%%%%%%%%%%%%%%%%%%%%%%%%%%%%%%%%%%%%%%%%
\section{Network Coding Aided Scheme}\label{S:DissNC}
Suppose BS is using linear random network coding. That means, in each transmission it sends a linear combination of the $M$ packets, where combination coefficients are uniformly chosen in finite field $\mathbb{F}_{2^q}$. By the same procedure we did for the feedback-based algorithm, we first derive the probability distribution and expectation of stopping time for a very simple case of $N=2$ and $m=1$. Then, we solve the problem in a more general case of $N=2$ and arbitrary $m$.

\subsection{$m=1$, $N=2$}
The BS sends a linear random combination of its $M$ packets in R2V phase. For example, in the first round, it sends the following packet to the first vehicle:

\begin{equation}\label{E:P1}
\sum_{i=1}^M\alpha_{1,i}x_i,\,\,\,\alpha_{1,i}\in\mathbb{F}_{2^q},
\end{equation}
where $x_i$'s are the BS information packets, as defined in Section \ref{S:I2V_mode}. The following combination is sent to the second vehicle:

\begin{equation}\label{E:P2}
\sum_{i=1}^M\alpha_{2,i}x_i,\,\,\,\alpha_{2,i}\in\mathbb{F}_{2^q}.
\end{equation}

At the end of the first round (after one R2V+V2V), both vehicles have the same information, i.e., they both have packets in Eq. \eqref{E:P1} and Eq. \eqref{E:P2}. The above can be written as $\mathbf{Ax}$ where $\mathbf{x}=[x_1\,x_2\,\ldots\, x_M]^T$ is an $M\times 1$ vector of BS packets and $\mathbf{A}$ is a $2\times M$ matrix containing NC coefficients. Clearly, at the end of second round two rows are added to matrix $\mathbf{A}$ and $\mathbf{Ax}$ is a $4\times 1$ vector representing 4 packets in each vehicle. In general at the end of each round, two rows are added to $\mathbf{A}$. Therefore, at the end of round $t$, matrix $\mathbf{A}$ has dimension $2t\times M$. Clearly, nodes are able to recover BS information whenever matrix $\mathbf{A}$ is invertible or equivalently when $rank(\mathbf{A})=M$. Thus, we can define the stopping time when NC is applied, $\mathcal{T}_{NC}$, as follows:

\begin{equation}\label{E:stoppingTime}
\mathcal{T}_{NC} = \min_t\{rank(A_{2t\times M})=M\}.
\end{equation}

Obviously $\mathcal{T}_{NC}\ge M/2$. As mentioned before, our goal is to calculate the probability distribution and expected value of $\mathcal{T}_{NC}$. The following lemma gives the probability of rank of matrix $\mathbf{A}_{t\times M}$ with random entries in a finite field.

\begin{lemma}
Let $\mathbf{A}_{t\times n}$ be a random matrix over finite field $\mathbb{F}_{2^q}$ such that each entry $a_{i,j}$ is picked uniformly from $\mathbb{F}_{2^q}$. suppose $t\ge n$. Then, probability of $\mathbf{A}$ having rank $n$ is given as follows:

\begin{eqnarray}\label{E:rank_prob}
P(rank(\mathbf{A}_{t\times n})=n) &=& \prod_{i=1}^n (1-{1\over q^{-(t-n+i)}}) \\ \nonumber
&\approx& 1-{1\over q^{-(t-n+1)}},
\end{eqnarray}

where the approximation is valid for sufficiently large $q$.
\end{lemma}
\emph{Proof}: See \cite{Migler}.
\begin{flushright}
$\blacktriangle$
\end{flushright}

Using the above result, an upper-bound on the probability of the stopping time can be calculated as given in following lemma. Its proof is omitted here for the sake of space limitation.

%SR Why call this I2V stopping time. There is only one stopping time, without subscript previously - be consistent.

%mhf I changed it to \mathcal{T}_NC to distinguish two cases

\begin{lemma}\label{Tm:Tprob}
Let $\mathcal{T}_{NC}$ be stopping time defined in Eq. \eqref{E:stoppingTime}. Then,
\begin{eqnarray}\label{E:PT}
P(\mathcal{T}_{NC}=t)\!\!\!\! &\le& \!\!\!\!(1\!-\!{1\over q})(1\!-\!\!P(\!rank(\!\mathbf{A}_{2t-1\times M})\!=\!M)) \\ \nonumber
\!\!\!\!&\le&\!\!\!\! (1-{1\over q}){1\over{q^{-(2t-M)}}},
\end{eqnarray}
where the last inequality is valid for large $q$.
\end{lemma}
\emph{Proof}: See \cite{hamed}.
\begin{flushright}
$\blacktriangle$
\end{flushright}
Finally, the following theorem gives an upper-bound for the expected value of the stopping time when $q$ is large enough.

\begin{theorem}\label{Tm:ET_I2V_NC}
Let $\mathcal{T}_{NC}$ be stopping time defined in \eqref{E:stoppingTime}. Suppose $q$ is large enough such that the approximation results in equations \eqref{E:rank_prob} and \eqref{E:PT} are valid. Then the expected value of stopping time is bounded from above with the following:
\begin{equation}
\mathbb{E}[\mathcal{T}_{NC}]\le{M\over 2}+{1\over q-1}.
\end{equation}
\end{theorem}
\begin{flushright}
$\blacktriangle$
\end{flushright}
%Clearly $M/2\le \mathbb{E}[\mathcal{T}_{NC}]< M/2+1$.
\subsection{$N=2$ and arbitrary $m$}
Results in previous section can be readily generalized to arbitrary $m$, number of packets transmitted from BS to vehicles in each R2V phase. Its proof is omitted here for the sake of space limitation.
\begin{theorem}\label{Tm:ET_R2V_NCGen}
Let $\mathcal{T}_{NC}$ be stopping time defined in \eqref{E:stoppingTime}. Suppose $q$ is large enough such that the approximation results in equations \eqref{E:rank_prob} and \eqref{E:PT} are valid. Then the expected value of stopping time is bounded from above with the following:
\begin{equation}
\mathbb{E}[\mathcal{T}_{NC}]\le {M\over {2m}}+{1\over q-1}
\end{equation}.
\end{theorem}
\begin{flushright}
$\blacktriangle$
\end{flushright}
\section{Evaluation Results}\label{S:Eval}
In this section, we present results from simulations conducted using MATLAB R2008b. We conduct simulations to confirm the analytical results in Section \ref{S:DissNoNC} and Section \ref{S:DissNC}, describing data dissemination in R2V phase with and without network coding.
%to demonstrate advantage of using network coding for collaborative downloading.

\begin{figure}
\psfig{figure=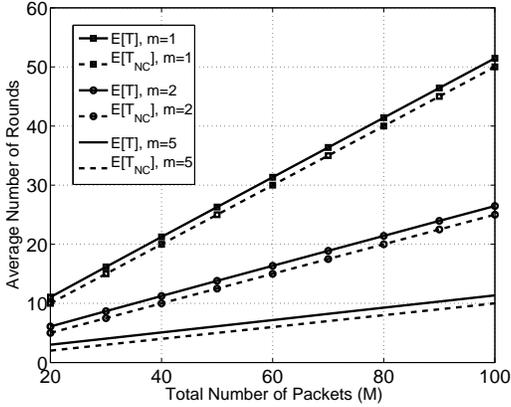,width=3.0in}
\caption{Average stopping time for $N=2$, $m=1,2,5$ and q=8}
\label{F:I2V_comp}
\end{figure}

Figure \ref{F:I2V_comp}  depicts average number of rounds needed to disseminate information from infrastructure to the vehicles. As explained before, we are assuming complete information exchange in V2V phase. In the simulation, the BS contains $M$ packets and in each round it sends $m$ packets to each vehicle separately. As we mentioned before, in this paper we limit ourselves to $N=2$ number of vehicles.

As one can see, in Figure \ref{F:I2V_comp}, $\mathcal{T}$ linearly increases with $M$, number of total information packets the BS possesses. There is very small benefit in using NC when the matric is number of rounds. In fact, for only two vehicles, the feedback-based algorithm needs almost one round more than NC aided scheme (note that NC almost achieves minimum number of rounds). Therefore, for dissemination delay, NC has small advantage compare to feedback-based algorithm. However, NC scheme removes any need for signaling. In NC aided scheme the BS keeps sending network coded data to the vehicles without any knowledge of what information they possess. On the other hand, for the feedback-based algorithm vehicles are required to communicate with BS before receiving any data. They need to inform the BS in what packets they are interested.
%On the other hand, uplink communication no longer is needed by using NC.

\section{Conclusion}\label{S:conclude}
In this paper, we consider amount of time needed (in R2V+V2V rounds) to distribute an object of $M$ packets to two vehicles in a VANET network. We show that using NC has the advantage of removing any need for signaling. In other words, to transmit the data from infrastructure to the vehicles there is no need for the BS to know which packets the vehicles already possess. This advantage of using NC comes with no cost. In fact, the data dissemination latency is slightly better when NC is applied. However, our analysis here is limited to two vehicles. Generalizing the result to arbitrary number of vehicles in the network is one of our future research area.

\bibliographystyle{IEEEtran}
\bibliography{IEEEabrv,overall}

\appendices

\section{Proofs of Theorems}
\emph{Proof of Lemma} \ref{Tm:P(X_i)}:

If we know $\mathcal{S}_{i-1}=s_i$, it means AP has $M-s_i$ new packets to send to the vehicles. Having that, The denominator equals all possible $m=1$ objects from $M-s_i$ chosen without replacement. The numerator represents all the possibilities such that there are exactly $x$ common among the $m$-set at each of the nodes.
\begin{flushright}
$\blacksquare$
\end{flushright}

%%%%%%%%%%%%%%%%%%%%%%%%%%%%%%%%%%%%%%%%%%%%%%%%%%%%%%%%%%%%%%%%%%%%%%%%%%%%%%%%%%%%%%%%%%%%%%%%%%%%%%%%%%%%

\emph{Proof of lemma} \ref{Tm:Tprob}

For matrix $\mathbf{A}_{m\times n}$, we define $span\{\mathbf{A}_{m\times n}\}$ as the space spanned by rows of matrix
$\mathbf{A}_{m\times n}$. By definition of stopping time $\mathcal{T}_{I2V}$ in Eq. \eqref{E:stoppingTime}, the event of $\mathcal{T}=t$ is equivalent of having $rank(\mathbf{A}_{2t\times M})=M$. Now, let $\mathbf{a}_{2t}$ be the last row of matrix $\mathbf{A}_{2t\times M}$ then we have:
\begin{eqnarray}\label{PT_I2V_proof}
P(\mathcal{T}_{I2V}=t) &=& P(\mathbf{A}_{2t\times M}=M) \\ \nonumber
             &=& P(rank(\mathbf{A}_{2t-1\times M})=M-1 \\ \nonumber
             && \,\,\, \text{ \& }\mathbf{a}_{2t} \not\in span\{\mathbf{A}_{2t-1\times M}\}) \\ \nonumber
             &=& P(rank(\mathbf{A}_{2t-1\times M})=M-1) \\ \nonumber
             && \,\,\, \cdot P(\mathbf{a}_{2t} \not\in span\{\mathbf{A}_{2t-1\times M}\} \\ \nonumber
             && \,\,\,\,\,\, |rank(\mathbf{A}_{2t-1\times M})=M-1).
\end{eqnarray}
Given the fact that rank of $\mathbf{A}_{2t-1\times M}$ is $M-1$, the probability of adding a row which is not independent of all the other rows can be bounded from below as follows:
\begin{eqnarray*}
&&P(\mathbf{a}_{2t}\in span\{\mathbf{A}_{2t-1\times M}\}|rank(\mathbf{A}_{2t-1\times M})=M-1) \\ \nonumber
&&\ge P(\mathbf{a}_{2t}=0|rank(\mathbf{A}_{2t-1\times M})=M-1) \\ \nonumber
&& ={1\over q}.
\end{eqnarray*}

Therefore, we have the following bound for having the last row$\mathbf{A}_{2t\times M}$ being independent of the rest of the rows:
\begin{eqnarray}\label{PT_I2V_proof1}
&&P(\mathbf{a}_{2t}\not\in span\{\mathbf{A}_{2t-1\times M}\} \\ \nonumber
&&\,\,\,\,\,\,|rank(\mathbf{A}_{2t-1\times M})=M-1)  \\ \nonumber
&&= 1 -P(\mathbf{a}_{2t}\in span\{\mathbf{A}_{2t-1\times M}\} \\ \nonumber
&&\,\,\,\,\,\,\,\,\, |rank(\mathbf{A}_{2t-1\times M})=M-1)  \\ \nonumber
&&\le 1-{1\over q}.
\end{eqnarray}

On the other hand:
\begin{eqnarray}\label{PT_I2V_proof2}
P(\mathbf{A}_{2t-1\times M}=M-1) \!\! &\le& \!\! P(\mathbf{A}_{2t-1\times M}\le M-1) \\ \nonumber
\!\!&\le& \!\!(1-P(\mathbf{A}_{2t-1\times M}=M)) \\ \nonumber
\!\! &=& \!\!{1\over {q^{-(2t-n)}}}.
\end{eqnarray}
By substituting Eq. \eqref{PT_I2V_proof1} and Eq. \eqref{PT_I2V_proof} in Eq. \eqref{PT_I2V_proof} the lemma is proven.

\begin{flushright}
$\blacksquare$
\end{flushright}

%%%%%%%%%%%%%%%%%%%%%%%%%%%%%%%%%%%%%%%%%%%%%%%%%%%%%%%%%%%%%%%%%%%%%%%%%%%%%%%%%%%%%%%%%%%%%%%%%%%%%%%%%%%%555
\emph{proof of Theorem }\ref{Tm:ET_I2V_NC}:

\begin{eqnarray}
E[\mathcal{T}_{NC}] &=& \sum_{t={M\over 2}}^M t\cdot P({\mathcal{T}_{NC}=t})\\ \nonumber
           &\le& \sum_{t={M\over 2}}^M t\cdot (1-{1\over q}){1\over q^{-(2t-M)}} \\ \nonumber
           &=& (1-{1\over q}) q^{-M} \sum_{t={M\over 2}}^M tq^{2t} \\ \nonumber
           &\approx& {M\over 2}+{1\over{q-1}}.
\end{eqnarray}

\begin{flushright}
$\blacksquare$
\end{flushright}

\end{document}